\documentclass[12pt,]{article}
 \usepackage{graphicx}
 \usepackage[cp1251]{inputenc}
\begin{document}

 \vbox to 2truemm{\hsize=165truemm{}}
 \centerline {\bf\large The influence of the bar on the chaotic dynamics of globular}
 \centerline {\bf\large  clusters in the central region of the Galaxy}
 \bigskip
 \bigskip
 \centerline{\bf
            A.T. Bajkova\footnote [1]{E-mail bajkova@gaoran,ru}, A.A. Smirnov, V.V. Bobylev
                        }
 \bigskip
 \centerline{\small \it Main Astronomical Observatory of the Russian Academy of Sciences, Pulkovo}
 \bigskip
 \bigskip
{\bf Abstract.} The paper is devoted to the analysis of the influence of the galactic bar on the nature of the orbital motion (chaotic or regular) of globular clusters in the central region of the Galaxy with a radius of 3.5 kpc, which are subject to the greatest influence of the bar. The sample includes 45 globular clusters. To form the 6D phase space required for integrating the orbits, the most accurate astrometric data to date from the Gaia satellite (Vasiliev, Baumgardt, 2021) were used, as well as new refined average distances (Baumgardt, Vasiliev, 2021). The orbits of the globular clusters were obtained both in an axisymmetric potential and in a potential including the bar. The following, most realistic, bar parameters were adopted: mass $10^{10} M_\odot$, semi-major axis length 5 kpc, bar axis rotation angle 25$^o$, angular rotation velocity 40 km s$^{-1}$ kpc$^{-1}$. The analysis of the chaoticity/regularity of the orbital motion in both potentials was carried out using one of the most effective methods, namely, the frequency method, which consists in calculating the drift of fundamental frequencies. As a result, the influence of the bar on the dynamics of each GC of the sample was assessed. It is established that 8 GCs changed regular dynamics to chaotic under the influence of the bar, and 9 GCs changed chaotic dynamics to regular one.

\medskip


\medskip

\section*{Introduction}

This work is a continuation of a series of works by the authors [1,2,3,4,5,6,7] devoted to the study of the orbital dynamics of globular clusters (GCs). Thus, in work [1] a catalog of orbits of 152 galactic globular clusters is presented based on the latest astrometric data from the Gaia satellite (Gaia EDR3) [8], as well as new refined average distances [9]. In work [2] an analysis was performed (based on the same data) of the influence of the galactic bar on the orbital motion of globular clusters in the central region of the Galaxy. For this task, 45 globular clusters in the central galactic region with a radius of 3.5 kpc were selected. A list of these GCs is given below in the table with the analysis results. The orbits of the globular clusters were obtained both in an axisymmetric potential and in a potential including a bar model in the form of a triaxial ellipsoid. In this case, the mass, angular velocity of rotation and the size of the bar were varied. A comparison of such orbital parameters as apocentric and pericentric distances, eccentricity and maximum distance from the galactic plane was made.

The second stage of the research aimed at studying the influence of the bar on the orbital motion of globular clusters was devoted to the problem of identifying objects captured by the bar using spectral dynamics methods [3,4,5,6].

The third stage of the research was devoted to the analysis of the regularity/chaotic nature of the orbits of all 45 selected GCs using various methods [7]. Namely, 1) methods for calculating maximum characteristic Lyapunov indices (MCLI) (in the classical version and in the version with renormalization of the "shadow" orbit corresponding to the perturbed initial phase points relative to the "reference" orbit with given initial phase points), 2) MEGNO, 3) Poincare sections, 4) a frequency method based on calculating fundamental frequencies, as well as 5) a visual assessment based on images of the "reference" and "shadow" orbits. In this case, the model of the bar was adopted as a model of an elongated triaxial ellipsoid with the most probable parameters known from the literature (see, for example, [10, 11]): mass $10^{10} M_\odot$, length of the major semiaxis of 5 kpc, angle of inclination to the galactic $X$ axis 25$^o$, angular rotation velocity 40 km s$^{-1}$ kpc$^{-1}$.

Since the GCs in the central region of the Galaxy are subject to the greatest influence from the elongated rotating bar, the question of the nature of the orbital motion of the GCs~-- regular or chaotic~--- is of great interest. For example, in [12] it is shown that
the main share of chaotic orbits should be precisely in the bar region.

This work is essentially a continuation of the third stage, devoted to the study of the chaotic dynamics of the selected GCs in the central region of the Galaxy. If in [7] we investigated the orbital dynamics of GCs only in a potential with a bar, then in this paper the task is to compare the orbital dynamics of GCs in an axisymmetric potential and in a non-axisymmetric potential in order to determine how the bar affects the degree of chaos of GC orbits. If in the previous paper we used several methods for analyzing the regularity/chaos of orbits, then in this paper we limit ourselves to using the most effective method, namely, the frequency method. For additional control of the obtained results, we also use the Poincare cross-section method, the result of which, as it turned out in [7], gives the highest correlation (about 96\%) with the results of using the frequency method.

The paper is structured as follows. The first section gives a brief description of the adopted potential models~--- an axisymmetric potential and a non-axisymmetric potential including a bar. The second section provides links to the astrometric data used, as well as to the method for forming the GC sample. The third section describes two methods for assessing the regularity/chaoticity of motion --- the Poincare section method and the frequency method. The fourth section analyzes the results obtained. The Conclusion formulates the main results of the work.

\section{Galactic Potential Model}

\subsection{Axisymmetric potential}

The axisymmetric gravitational potential of the Galaxy, traditionally used by us (see, for example, [1]) for integrating the GC orbits, is represented as the sum of three components~--- the central spherical bulge $\Phi_b(r)$, the disk $\Phi_d(R,Z)$ and the massive spherical halo of dark matter $\Phi_h(r)$:
 \begin{equation}
 \begin{array}{lll}
  \Phi(R,Z)=\Phi_b(r)+\Phi_d(R,Z)+\Phi_h(r).
 \label{pot}
 \end{array}
 \end{equation}
Here we use a cylindrical coordinate system ($R,\psi,Z$) with the origin at the center of the Galaxy. In a rectangular coordinate system $(X,Y,Z)$ with the origin at the center of the Galaxy, the distance to a star (spherical radius) will be $r^2=X^2+Y^2+Z^2=R^2+Z^2$, with the $X$ axis directed from the Sun to the galactic center, the $Y$ axis perpendicular to the $X$ axis in the direction of the Galaxy's rotation, and the $Z$ axis perpendicular to the galactic plane $(X,Y)$ toward the north galactic pole. The gravitational potential is expressed in units of 100 km$^2$ s$^{-2}$, distances~--- in kpc, masses~--- in units of the galactic mass $M_{gal}=2.325\times 10^7 M_\odot$,
corresponding to the gravitational constant $G=1$.

The axisymmetric potentials of the bulge $\Phi_b(r(R,Z))$ and the disk $\Phi_d(r(R,Z))$ are represented in the form proposed by Miyamoto, Nagai~[13]:
 \begin{equation}
  \Phi_b(r)=-\frac{M_b}{(r^2+b_b^2)^{1/2}},
  \label{bulge}
 \end{equation}
 \begin{equation}
 \Phi_d(R,Z)=-\frac{M_d}{\Biggl[R^2+\Bigl(a_d+\sqrt{Z^2+b_d^2}\Bigr)^2\Biggr]^{1/2}},
 \label{disk}
\end{equation}
where $M_b, M_d$~ are masses of components, $b_b, a_d, b_d$~ are scale parameters of components in kiloparsecs. The halo component (NFW) is represented according to the work~[14]:
 \begin{equation}
  \Phi_h(r)=-\frac{M_h}{r} \ln {\Biggl(1+\frac{r}{a_h}\Biggr)}.
 \label{halo-III}
 \end{equation}
Table~1 presents the values of the parameters jf the galactic potential model (\ref{bulge})---(\ref{halo-III}), which were found by Bajkova and Bobylev~[15] using the Galactic rotation curve [16], constructed based on objects located at distances $R$ up to $\sim200$~kpc. Note that when constructing this Galactic rotation curve, the following values of the local parameters: $R_\odot=8.3$~kpc and $V_\odot=244$~km s$^{-1}$ were used. In work~[5], model (\ref{bulge})---(\ref{halo-III}) is designated as model~III. The adopted potential model is the best among the six models considered in work [17], since it provided the smallest discrepancy between the data and the model rotation curve.
\begin{figure*}
{\begin{center}

\includegraphics[width=0.4\textwidth,angle=-90]{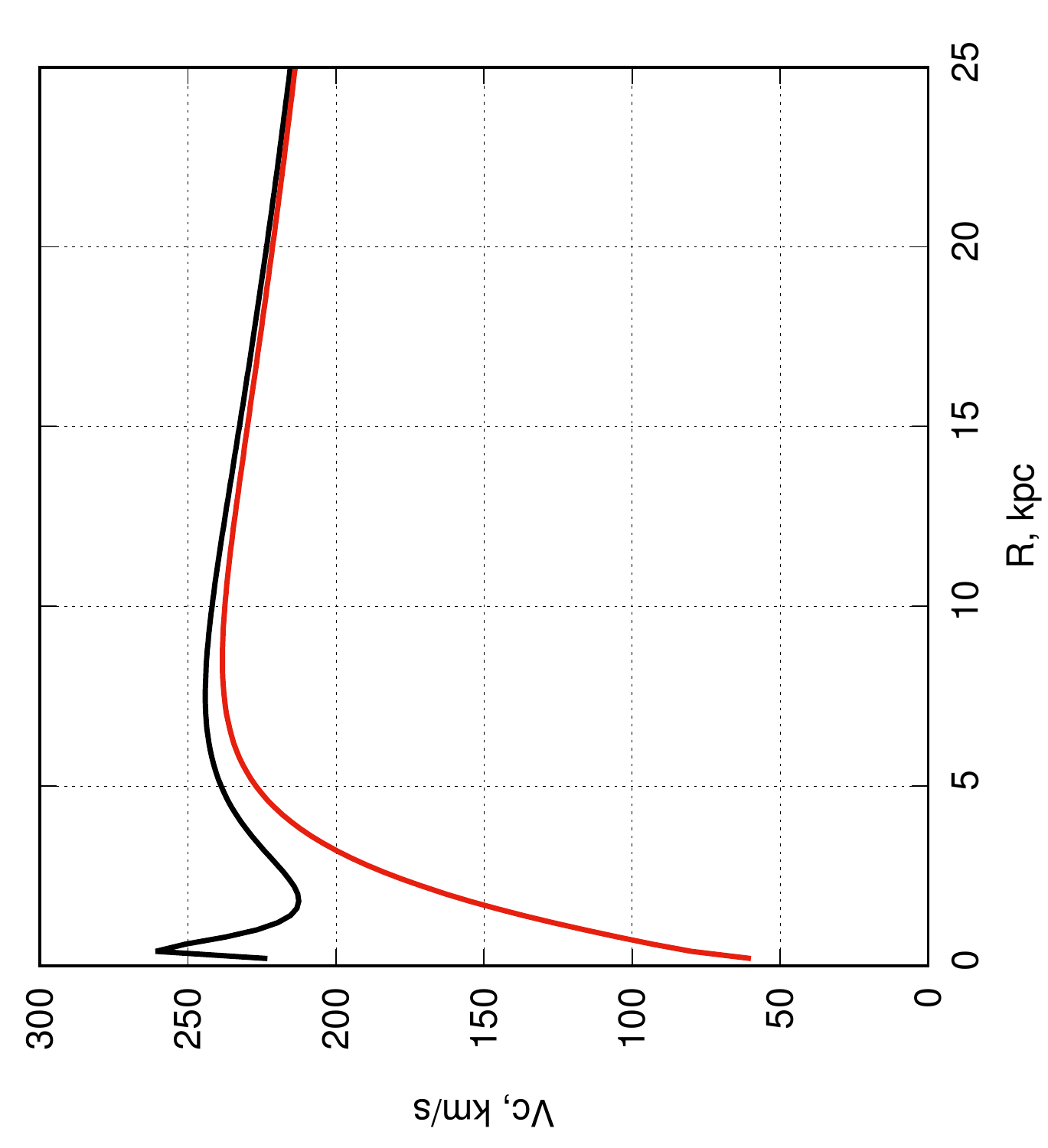}

\bigskip

\caption{\small Rotation curve of the Galaxy with an axisymmetric potential without a bar (black line) and a non-axisymmetric potential including a bar (red line).}
\label{fcomp}
\end{center}}
\end{figure*}

\subsection{Bar model}

The triaxial ellipsoid model was chosen as the central bar potential~[10]:
\begin{equation}
  \Phi_{bar} = -\frac{M_{bar}}{(q_b^2+X^2+[Ya/b]^2+[Za/c]^2)^{1/2}},
\label{bar}
\end{equation}
where $X=R\cos\vartheta, Y=R\sin\vartheta$, $a, b, c$~are three
semi-axes of the bar, $q_b$~is scale parameter of the bar (length of the largest semi-axis of the bar);
$\vartheta=\theta-\Omega_{b}t-\theta_{b}$, $tg(\theta)=Y/X$,
$\Omega_{b}$~is circular velocity of the bar, $t$~is integration time, $\theta_{b}$~is orientation angle of the bar
relative to the galactic axes $X,Y$ measured from the line
connecting the Sun and the center of the Galaxy (axis $X$) to the major axis of the bar in the direction of rotation of the Galaxy.

Based on information in numerous literature, in particular, in [10], the following were used as bar parameters: $M_{bar}=430 M_{gal}$, $\Omega_{b}=40$~km s$^{-1}$ kpc$^{-1}$, $q_b=5$ kpc, $\theta_{b}=25^o$. The adopted bar parameters are listed in Table~1.

 {\begin{table}[t]                                    
 \caption[]
 {\small\baselineskip=1.0ex
Parameters of the galactic potential model, $M_{gal}=2.325\times 10^7 M_\odot$
  }
 \label{t:model-III}
 \begin{center}\begin{tabular}{|c|r|}\hline
 $M_b$ &   443 M$_{gal}$ \\
 $M_d$ &  2798 M$_{gal}$ \\
 $M_h$ & 12474 M$_{gal}$ \\
 $b_b$ & 0.2672 kpc  \\
 $a_d$ &   4.40 kpc  \\
 $b_d$ & 0.3084 kpc  \\
 $a_h$ &    7.7 kpc  \\
\hline\hline
 $M_{bar}$ & 430 M$_{gal}$ \\
 $\Omega_b$ & 40 km s$^{-1}$ kpc$^{-1}$ \\
 $q_b$     &  5.0 kpc  \\
 $\theta_{b}$ &  $25^o$   \\\hline
 $a/b$ & 2.38  \\
 $a/c$ & 3.03  \\
    \hline
 \end{tabular}\end{center}\end{table}}

To integrate the equations of motion, we used the fourth-order Runge-Kutta algorithm.

The value of the peculiar velocity of the Sun relative to the local standard of rest was taken to be $(u_\odot,v_\odot,w_\odot)=(11.1,12.2,7.3)\pm(0.7,0.5,0.4)$~km s$^{-1}$
according to the work~[18]. The elevation of the Sun above the plane of the Galaxy was taken to be 16 pc in accordance with the work [19].

For comparison, Fig.~1 shows the obtained model rotation curves: an axisymmetric potential (black line) and a potential with a bar (red line).

\section{Data}

The data on the proper motions of GCs are taken from the new catalog by Vasiliev and Baumgardt, 2021 [8], compiled on the basis of Gaia EDR3 observations. The GC coordinates and radial velocities are taken from [21]. The average values of distances to globular clusters are taken from Baumgardt and Vasiliev, 2021 [9]. A comparative analysis of the new data on proper motions and distances with previous versions of the catalogs is given, for example, in [1].

The catalog of GCs at our disposal [1] contains 152 objects.
The selection of globular clusters from this set, belonging to the bulge/bar region, was made in accordance with a purely geometric criterion, considered in [21], and also used by us in [22]. It is very simple and consists of selecting GCs, the apocentric distance of whose orbits does not exceed the bulge radius, which is usually taken to be 3.5 kpc. The orbits are calculated in an axisymmetric potential. The full list of 45 objects in our sample is listed in Table~2, which provides the results of the analysis of the GC orbital randomness/regularity (the first column gives the GC serial number, the second column gives the GC name).

\section{Methods of analysis of regularity/chaoticity of GC orbital dynamics}

\subsection{Poincare sections}

One of the methods for determining the nature of the motion (regular or chaotic) is the analysis of Poincare sections [23]. The algorithm we used to construct the mappings is as follows:

1. We consider the phase space $(X,Y,V_x,V_y)$.

2. We eliminate $V_y$ using the conservation law of the generalized energy integral (Jacobi integral) and move to the space $(X,Y,V_x)$.

3. We define the plane $Y=0$, the intersection points with the orbit are designated on the plane $(X,V_x)$. We take only those points where $V_y>0$.

Similarly, the phase space $(Y,Z,V_y,V_z)$ or $(R,Z,V_R,V_z)$ can be considered. Then the Poincare sections will be reflected on the plane $(Y,V_y)$ or $(R,V_R)$, respectively.

If the intersection points of the plane add up to a continuous smooth line (or several separated lines), then the motion is considered regular. In the case of chaotic motion, instead of being located on a smooth curve, the points fill a two-dimensional region of phase space, and sometimes the effect of sticking points to the boundaries of islands corresponding to ordered motion occurs [23].

 It is important to note that for non-axisymmetric potential models, which include one of the potentials considered in this paper, including a rotating central bar, the Poincare sections have a more complex structure than in the case of an axisymmetric model. If in the case of axisymmetric models for regular orbits the Poincare sections, as a rule, represent a straight line, then in the case of non-axisymmetric models for many orbits more complex patterns are obtained. It would be incorrect to call such orbits chaotic, since obvious patterns are observed in the arrangement of points, but they may no longer form a single line. Thus, the problem of dividing orbits into regular and chaotic based on Poincare sections is noticeably more complicated and is not devoid of subjectivity. Therefore, it is of great importance to involve, along with Poincare sections, other methods of analysis and to make a decision on the nature of the motion of objects based on the results of using several independent approaches. In this case, we use the frequency method described in the next paragraph, which was the main one in deciding on the regularity (R) or chaos (C) of each GC orbit from our sample.

 \subsection{Frequency method}

The described method of studying the regularity/chaoticity of orbits is associated with the use of orbital frequencies [24,25] (see Section 3.1 in the last paper). The authors of these works showed that it is possible to measure the stochasticity of an orbit based on the shift of fundamental frequencies determined over two consecutive time intervals. For each frequency component $f_i$, a parameter called the frequency drift is calculated:
\begin{equation}
\label{freq}
\lg(\Delta f_i)=\lg|\frac{\Omega_i(t_1)-\Omega_i(t_2)}{\Omega_i(t_1)}|,
\end{equation}
where $i$ defines the frequency component in Cartesian coordinates (i.e. $\lg(\Delta f_x), \lg(\Delta f_y)$ and $\lg(\Delta f_z)$). Then the largest value of these three frequency drift parameters $\lg(\Delta f_x)$ is assigned to the frequency drift parameter $\lg(\Delta f)$. The higher the value of $\lg(\Delta f)$, the more chaotic the orbit. However, as shown in [25], the accuracy of frequency analysis requires at least 20 oscillation periods to avoid classification errors.

We calculated the frequency drift parameter $\lg(\Delta f)$ for all 45 GCs in both potentials, which was used to determine the nature of their motion --- (R) or (C). The $x(t_n), y(t_n), z(t_n)$ series were determined over the time interval [0, 120] billion years. The first power spectrum of each GC was calculated over the time interval [0, 60] billion years, the second --- over [60, 120] billion years. Then the frequency drift parameters were calculated for each time series $x(t_n), y(t_n), z(t_n)$ using the formula (\ref{freq}). The largest value of them was taken as the frequency drift parameter $\lg(\Delta f)$. In the case of the coincidence of fundamental frequencies $\Omega_i(t_1)=\Omega_i(t_2)$, we artificially assumed the frequency drift parameter to be equal to $-4$.

\section{Results}

The obtained results of applying the Poincare sections method and the frequency method to determine the nature of the orbital dynamics of the GC in the center of the Galaxy are reflected in Table~2 and Fig.~2 (see the figure caption).

The solutions on the regularity (R) or chaos (C) of the orbital motion of all 45 GCs, obtained on the basis of the analysis of the Poincare sections on the plane $(X,V_x)$ in the axisymmetric and non-axisymmetric potential with a bar, are given in Table~2 in the 3rd and 4th columns, respectively. A graphical illustration of the Poincare sections method is given in Fig.~2 in the 2nd and 5th vertical rows of the panels for the axisymmetric and non-axisymmetric potentials, respectively.

The results of calculating the frequency drift parameter for all 45 GCs, both in the axisymmetric potential and in the potential with a bar, are given in Table~2 in columns 5 and 6, respectively.
A graphical illustration of the frequency method, namely, the results of calculating the power spectra, is shown in Fig.~2 in the 3rd and 6th vertical rows of the panels for the axisymmetric and non-axisymmetric potentials, respectively.

The decision on the nature of the motion (regular (R) or chaotic (C)) when using the frequency method was made in accordance with the recommendations set out in [7] (only with a threshold value of the frequency drift parameter equal to $-2.14$). A smaller value than $-2.14$ corresponds to regular orbits, a larger value corresponds to chaotic ones, with the exception of two GCs: Terzan 3 and NGC 6316, for which $\lg(\Delta f)\approx -2$, which are classified according to the results of visual analysis of the power spectra (Fig.~2) as GCs with regular orbits, although they show a weak degree of chaos. This was also done in the previous work [7].

The difference in the frequency drift parameters in the axisymmetric and non-axisymmetric potentials is given in the 7th column of Table~2. A negative difference indicates an increase in the degree of chaos of the orbital motion in the potential with a bar, while a positive difference indicates a decrease in the degree of chaos.

The final decision on the nature of the motion and its change depending on the presence of a central bar in the Galaxy was made based on the frequency method and is reflected in the 8th column of Table~2. It should be noted that the correlation between the results of the Poincare cross-section method and the frequency method for both potentials was 96\%, which coincides with the result we obtained in [7] for a non-axisymmetric potential with a bar.

As the analysis of the last column of Table~2 shows, the inclusion of a bar in the axisymmetric potential noticeably affected the orbital dynamics of the GC in our sample. Thus, in 8 GCs (NGC 6144, Ngc 6273, NGC 6342, NGC 6355, NGC 6558, NGC 6256, NGC 6304, NGC 6388) the regular dynamics was replaced by chaotic (designated as $R\rightarrow C$); in 9 GCs (Terzan 4, Liller1, NGC 6380, Terzan 5, NGC 6440, Terzan 6, Terzan 9, NGC 6624, NGC 6637) the chaotic dynamics was replaced by regular ($C\rightarrow R$); In 17 GCs the degree of chaos changed slightly, which did not lead to a change in the status of regular or chaotic motion (in 12 it increased: $C\uparrow$), and in 5 it decreased: ($C\downarrow$)), and only in 11 GCs (NGC 6266, Terzan 1, NGC 6522, NGC 6717, NGC 6723, Pismis 26, NGC 6569, NGC 6540, NGC 6171, NGC 6539, NGC 6553) the regular dynamics remained without any changes ($=$).

As the analysis of Fig.~2 shows, the greatest influence of the bar was experienced by GCs with orbits elongated in the radial direction (large eccentricities and small pericentric distances and high radial velocities), which coincides with the conclusions of our previous work [7].


\begin{figure*}
{\begin{center}
       \includegraphics[width=1\textwidth,angle=0]{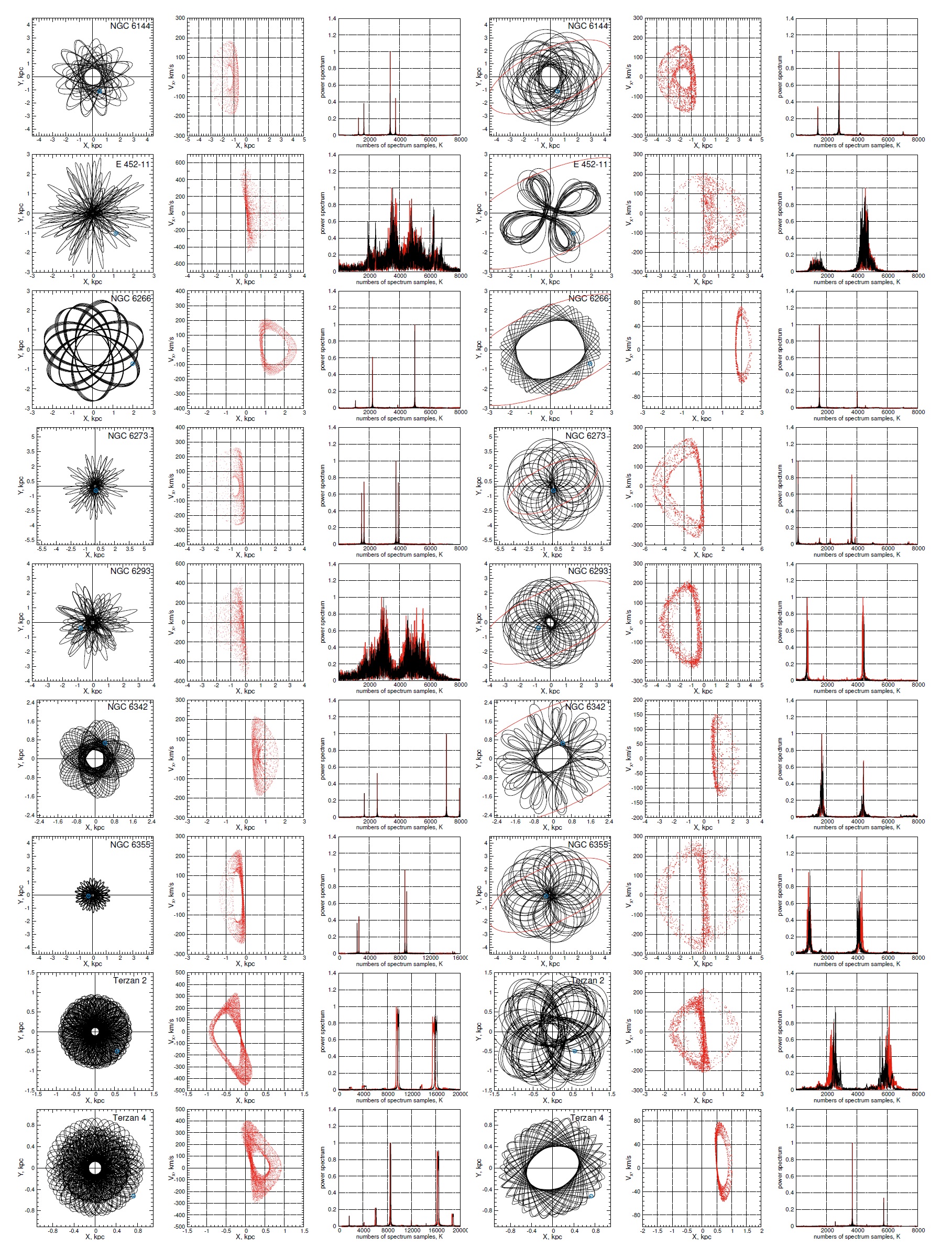}
    \caption{\small Graphic illustration of the orbital dynamics of 45 GCs in an axisymmetric (left triple of vertical rows of panels) and a barred (right triple of vertical rows of panels) potentials. From left to right for each potential are shown the projections of the orbits onto the galactic plane $(X,Y)$ (in the second case they are shown in the rotating bar system, and the red lines are the bar sections); Poincare sections on the $(X,V_x)$ plane; power spectra of time sequences for the frequency method (red color refers to the first half of the sequence, black --- to the second.}
    \label{fig:f1}
\end{center}}
\end{figure*}

\begin{figure*}
{\begin{center}
       \includegraphics[width=1\textwidth,angle=0]{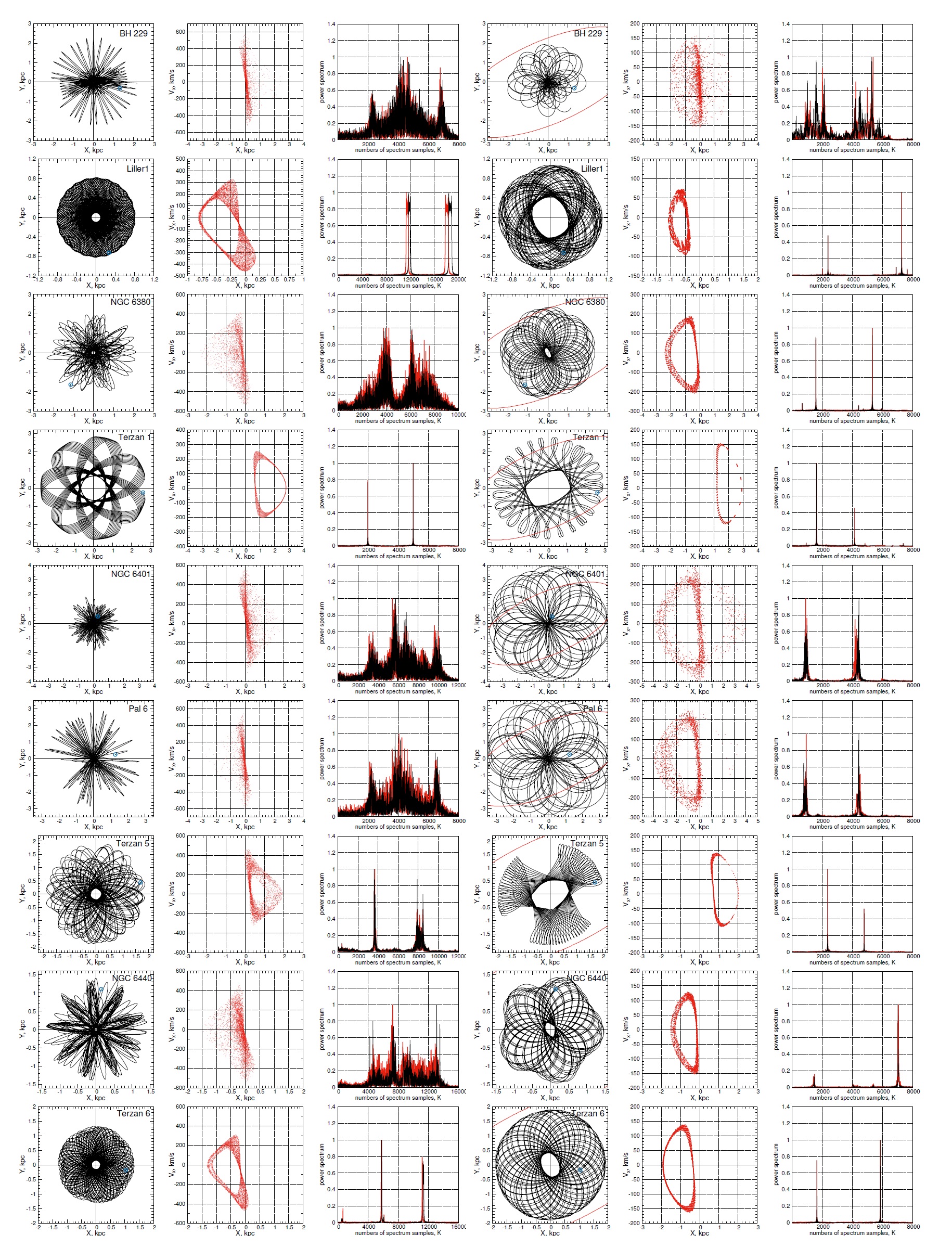}
    \centerline{Figure~2. Continuation.}
    \label{fig:f1}
\end{center}}
\end{figure*}

\begin{figure*}
{\begin{center}
       \includegraphics[width=1\textwidth,angle=0]{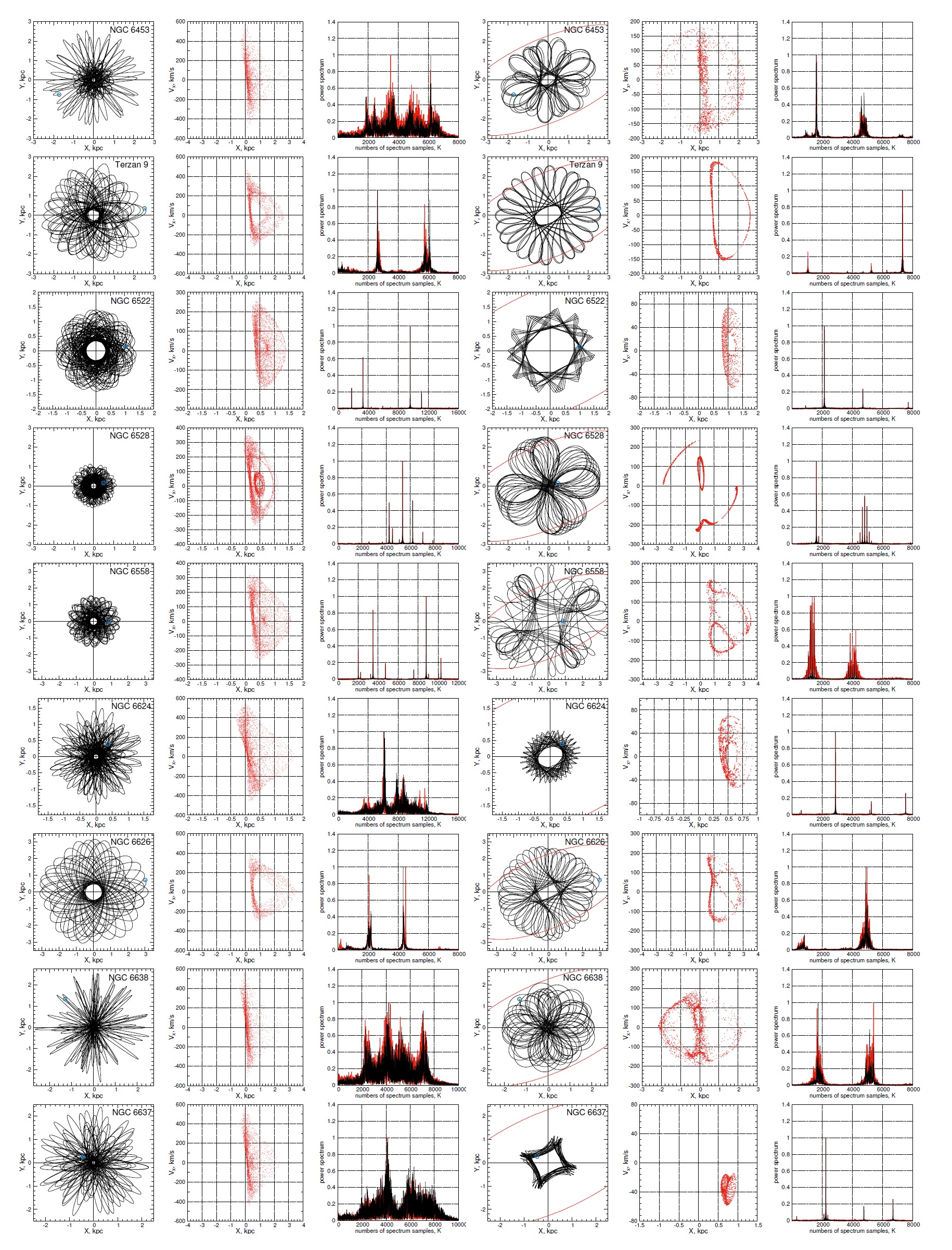}
    \centerline{Figure~2. Continuation.}
    \label{fig:f1}
\end{center}}
\end{figure*}

\begin{figure*}
{\begin{center}
       \includegraphics[width=1\textwidth,angle=0]{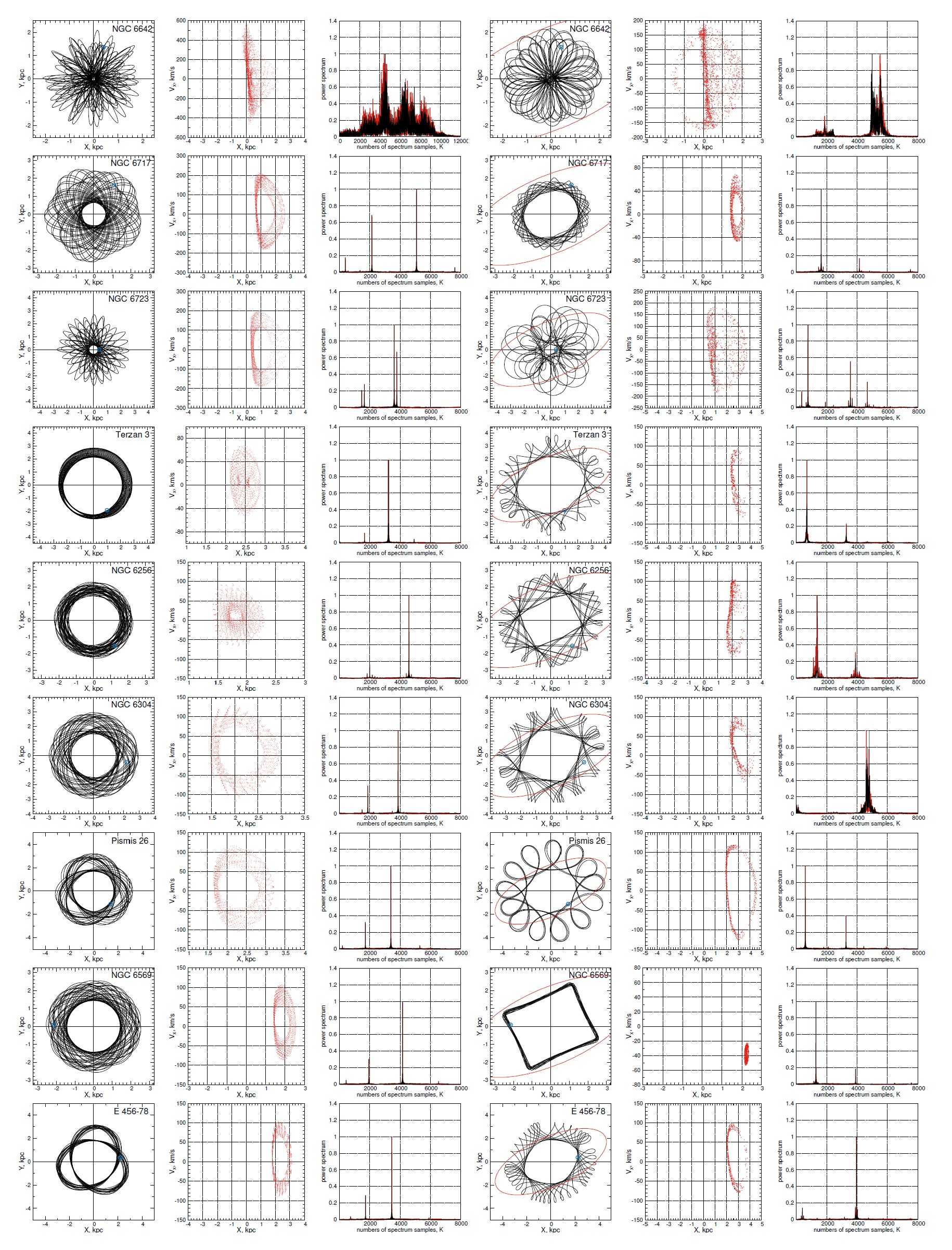}
    \centerline{Figure~2. Continuation.}
    \label{fig:f1}
\end{center}}
\end{figure*}

\begin{figure*}
{\begin{center}
       \includegraphics[width=1\textwidth,angle=0]{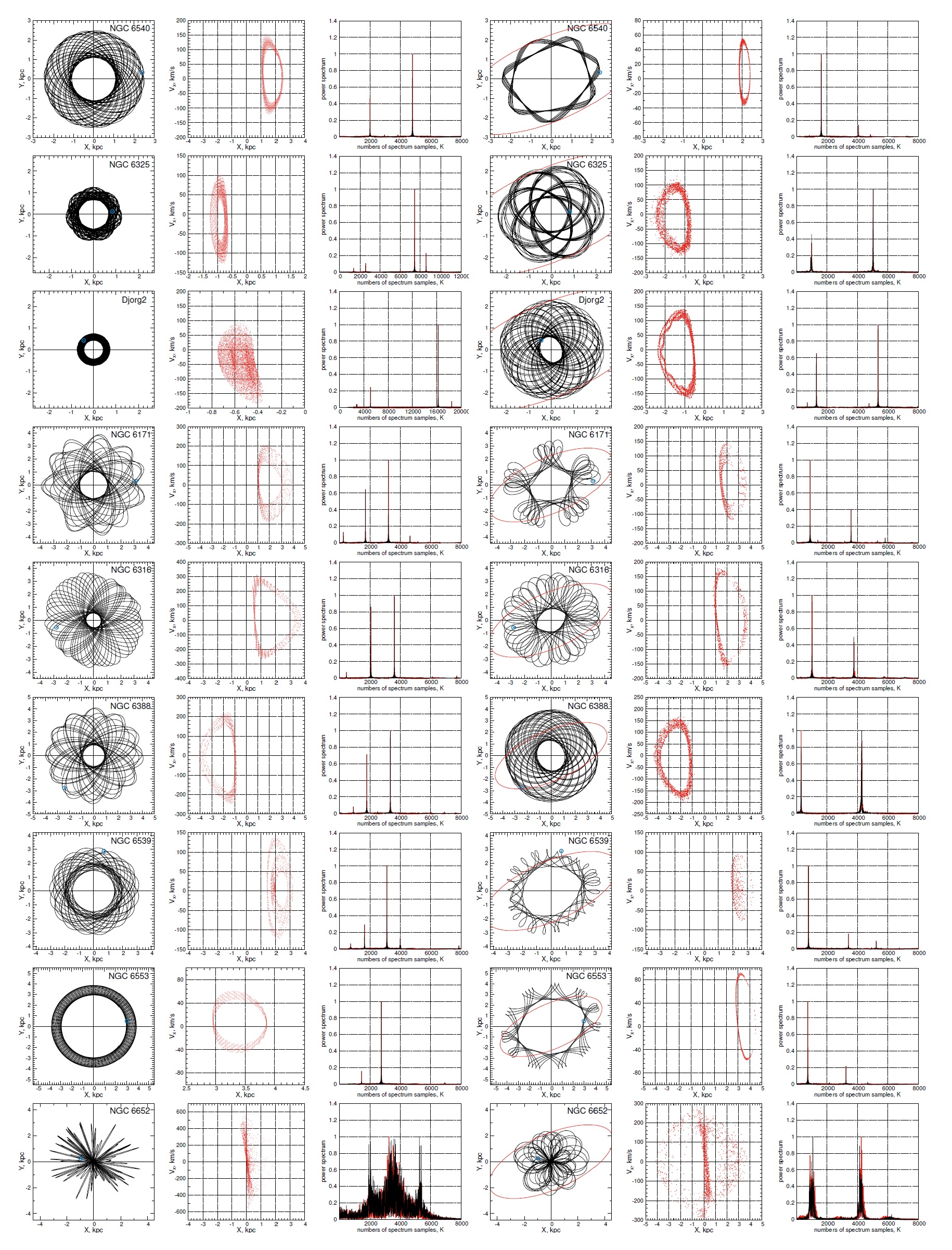}
    \centerline{Figure~2. Continuation.}
    \label{fig:f1}
\end{center}}
\end{figure*}

\section{Conclusion}

The paper addresses the problem of studying the influence of the galactic bar on the orbital dynamics of globular clusters in the central region of the Galaxy with a radius of 3.5 kpc. To solve this problem, orbits were constructed in both axisymmetric and non-axisymmetric potentials, including the bar. The following, most realistic parameters of the bar model in the form of a triaxial ellipsoid were adopted [10]: mass $10^{10} M_\odot$, length of the major semi-axis 5 kpc, angle of rotation of the bar axis 25$^o$, angular velocity of rotation 40 km s$^{-1}$ kpc$^{-1}$. A sample of 45 globular clusters, previously formed by us in paper [2], was used. To integrate the orbits, the most accurate astrometric data to date from the Gaia satellite (EDR3) [8] were used, as well as new refined average distances to globular clusters [9].

For the analysis and final decision on the chaotic/regular orbital motion of globular clusters in both potentials, one of the most effective methods was used, namely, the frequency method, which consists of calculating the drift of fundamental frequencies. For control, the Poincare cross-section method was also used. The correlation between these two methods was 96\%. As a result, the influence of the bar on the dynamics of each GC in our sample was assessed. 8 GCs (NGC 6144, NGC 6273, NGC 6342, NGC 6355, NGC 6558, NGC 6256, NGC 6304, NGC 6388) were identified that changed their regular dynamics to chaotic under the influence of the bar, and 9 GCs (Terzan 4, Liller1, NGC 6380, Terzan5, NGC 6440, Terzan 6, Terzan 9, NGC 6624, NGC 6637) changed their chaotic dynamics to regular. In 11 GCs (NGC 6266, Terzan 1, NGC 6522, NGC 6717, NGC 6723, Pismis 26, NGC 6569, NGC 6540, NGC 6171, NGC 6539, NGC 6553) the regular dynamics remained absolutely unchanged. In the remaining GCs the orbital dynamics underwent minor changes that did not lead to a change in the status of regular or chaotic motion.

As the analysis of the obtained graphic material (Fig.~2) showed, the GCs with elongated radial orbits (large eccentricities and small pericentric distances and high radial velocities) were subjected to the greatest influence of the bar, which coincides with the conclusions of our previous work [7].

\section*{Acknowledgements}
The authors are grateful to the reviewer for a number of useful comments that allowed us to eliminate inaccuracies in the classification of GC orbits and, thus, significantly improve the article.

 {\begin{table*}[t]                                    
 \caption[]
 {\baselineskip=1.0ex
The influence of the bar on the dynamics of the orbits of 45 GCs.
  }
 \label{t:f}
 {\scriptsize\begin{center}\begin{tabular}{|r|l|c|c||c|c|c||c|}\hline
  &      &Poincare sections&Poincare sections&Frequency drift&Frequency drift&Difference         &The nature   \\
 N&Names &in axisymmetric  &in potential     &in axisymmetric&in potential   &in drift of        &of the change\\
  &of GCs&potential        &with a bar       &potential      &with a bar     &frequencies$\Delta$&in dynamics  \\\hline
 1  &NGC6144 & (R) & (C) & -4.00 (R)   &-2.08 (C)      &  -1.91       &R$\rightarrow$C    \\\hline
 2  &E452-11 & (C) & (C) & -1.70 (C)   &-1.37 (C)      &  -0.32       &C$\uparrow$    \\\hline
 3  &NGC6266 & (R) & (R) & -4.00 (R)   &-4.00 (R)      &   0.00       &  $=$          \\\hline
 4  &NGC6273 & (R) & (C) & -4.00 (R)   &-1.77 (C)      &  -2.22       &R$\rightarrow$C\\\hline
 5  &NGC6293 & (C) & (C) & -1.34 (C)   &-0.07 (C)      &  -1.27       &C$\uparrow$    \\\hline
 6  &NGC6342 & (R) & (C) & -4.00 (R)   &-2.14 (C)      &  -1.85       &R$\rightarrow$C\\\hline
 7  &NGC6355 & (R) & (C) & -4.00 (R)   &-0.10 (C)      &  -3.89       &R$\rightarrow$C\\\hline
 8  &Terzan2 & (R) & (C) & -1.61 (C)   &-0.23 (C)      &  -1.37       &C$\uparrow$    \\\hline
 9  &Terzan4 & (C) & (R) & -1.97 (C)   &-4.00 (R)      &   2.02       &C$\rightarrow$R  \\\hline
10  &BH229   & (C) & (C) & -1.00 (C)   &-1.81 (C)      &   0.80       &C$\downarrow$  \\\hline
11  &Liller1 & (C) & (R) & -1.49 (C)   &-4.00 (R)      &   2.50       &C$\rightarrow$R\\\hline
12  &NGC6380 & (C) & (R) & -0.41 (C)   &-3.72 (R)      &   3.30       &C$\rightarrow$R\\\hline
13  &Terzan1 & (R) & (R) & -4.00 (R)   &-4.00 (R)      &   0.00       &  $=$          \\\hline
14  &NGC6401 & (C) & (C) & -1.26 (C)   &-0.09 (C)      &  -1.16       &C$\uparrow$    \\\hline
15  &Pal6    & (C) & (C) & -0.42 (C)   &-0.10 (C)      &  -0.32       &C$\uparrow$    \\\hline
16  &Terzan5 & (C) & (R) & -1.78 (C)   &-4.00 (R)      &   2.21       &C$\rightarrow$R\\\hline
17  &NGC6440 & (C) & (R) & -0.35 (C)   &-2.26 (R)      &   1.91       &C$\rightarrow$R\\\hline
18  &Terzan6 & (R) & (R) & -0.08 (C)   &-4.00 (R)      &   3.91       &C$\rightarrow$R\\\hline
19  &NGC6453 & (C) & (C) & -0.36 (C)   &-1.92 (C)      &   1.56       &C$\downarrow$  \\\hline
20  &Terzan9 & (C) & (R) & -0.00 (C)   &-3.86 (R)      &   3.86       &C$\rightarrow$R\\\hline
21  &NGC6522 & (R) & (R) & -3.98 (R)   &-4.00 (R)      &   0.01       &   $\approx$   \\\hline
22  &NGC6528 & (R) & (R) & -2.71 (R)   &-4.00 (R)      &   1.28       &C$\downarrow$  \\\hline
23  &NGC6558 & (R) & (C) & -3.09 (R)   &-1.03 (C)      &  -2.06       &R$\rightarrow$C\\\hline
24  &NGC6624 & (C) & (R) & -2.11 (C)   &-4.00 (R)      &   1.88       &C$\rightarrow$R\\\hline
25  &NGC6626 & (C) & (C) & -0.00 (C)   &-1.78 (C)      &   1.78       &C$\downarrow$  \\\hline
26  &NGC6638 & (C) & (C) & -1.49 (C)   &-0.16 (C)      &  -1.32       &C$\uparrow$    \\\hline
27  &NGC6637 & (C) & (R) & -1.68 (C)   &-4.00 (R)      &   2.31       &C$\rightarrow$R\\\hline
28  &NGC6642 & (C) & (C) & -1.66 (C)   &-1.01 (C)      &  -0.65       &C$\uparrow$    \\\hline
29  &NGC6717 & (R) & (R) & -4.00 (R)   &-4.00 (R)      &   0.00       &  $=$          \\\hline
30  &NGC6723 & (R) & (R) & -4.00 (R)   &-4.00 (R)      &   0.00       &  $=$          \\\hline
31  &Terzan3 & (R) & (R) & -4.00 (R)   &-1.89 (R)      &  -2.10       &C$\uparrow$    \\\hline
32  &NGC6256 & (R) & (C) & -4.00 (R)   &-1.93 (C)      &  -2.06       &R$\rightarrow$C\\\hline
33  &NGC6304 & (R) & (C) & -4.00 (R)   &-1.38 (C)      &  -2.61       &R$\rightarrow$C\\\hline
34  &Pismis26 & (R) & (R) & -4.00 (R)   &-4.00 (R)      &   0.00       &  $=$          \\\hline
35  &NGC6569 & (R) & (R) & -4.00 (R)   &-4.00 (R)      &   0.00       &  $=$          \\\hline
36  &E456-78 & (R) & (R) & -4.00 (R)   &-3.59 (R)      &  -0.40       &C$\uparrow$    \\\hline
37  &NGC6540 & (R) & (R) & -4.00 (R)   &-4.00 (R)      &   0.00       &  $=$          \\\hline
38  &NGC6325 & (R) & (C) & -4.00 (R)   &-3.22 (R)      &  -0.77       &C$\uparrow$    \\\hline
39  &Djorg2  & (R) & (R) & -3.90 (R)   &-4.00 (R)      &   0.09       &C$\downarrow$  \\\hline
40  &NGC6171 & (R) & (R) & -4.00 (R)   &-4.00 (R)      &   0.00       &  $=$          \\\hline
41  &NGC6316 & (R) & (R) & -4.00 (R)   &-1.96 (R)      &  -2.03       &C$\uparrow$    \\\hline
42  &NGC6388 & (R) & (C) & -4.00 (R)   &-0.03 (C)      &  -3.97       &R$\rightarrow$C\\\hline
43  &NGC6539 & (R) & (R) & -4.00 (R)   &-4.00 (R)      &   0.00       &  $=$          \\\hline
44  &NGC6553 & (R) & (R) & -4.00 (R)   &-4.00 (R)      &   0.00       &  $=$          \\\hline
45  &NGC6652 & (C) & (C) & -0.39 (C)   &-0.12 (C)      &  -0.26       &C$\uparrow$    \\\hline
 \end{tabular}\end{center}}\end{table*}}

\bigskip{\bf\large References}\bigskip {\small

1~A. T. Bajkova and V. V. Bobylev. A New Catalog of orbits of 152 Globular Clusters from Gaia EDR3. // Publications of the Pulkovo Observatory {\bf 227}, P. 1--15 (2022) DOI:10.31725/0367-7966-2022-227-2, arXiv: 2212.00739.

\medskip

2~A. T. Bajkova, A. A. Smirnov, and V. V. Bobylev. Globular clusters in the central region of the Milky Way galaxy I. Bar influence on the orbit parameters according to Gaia EDR3. // Publications of the Pulkovo Observatory {\bf 228}, P. 1--31 (2023) DOI:10.31725/0367-7966-2023-228-1, arXiv: 2305.05012.

\medskip

3~A. T. Bajkova, A. A. Smirnov, and V. V. Bobylev. Globular clusters in the central region of the Milky Way galaxy. II. Frequency
analysis of orbits built from Gaia ED3 data. // Publications of the Pulkovo Observatory {\bf 229}, P. 1--12 (2023) DOI:10.31725/0367-7966-2023-229-1.

\medskip

4.~A .A. Smirnov, A. T. Bajkova, V. V. Bobylev. Globular clusters captured by the Milky Way’s bar. // Publications of the Pulkovo Observatory {\bf 228}, P. 157--165 (2023) DOI:10.31725/0367-7966-2023-228-12.

\medskip

5.~A. T. Bajkova, A. A. Smirnov, and V. V. Bobylev.The Influence of the Bar on the Dynamics of Globular Clusters in the Central Region of the Milky Way. Frequency Analysis of Orbits According to Gaia EDR3 Data. // Astrophysical Bulletin {\bf 78}, Issue 4, P. 499--513 (2023), arXiv: 2311.14789.

\medskip

6.~A. A. Smirnov, A. T. Bajkova, V. V. Bobylev.  Globular clusters and bar: captured or not captured? // Monthly Notices of the Royal Astronomical Society {\bf 528}, Issue 2, P. 1422--1437 (2024), arXiv: 2310.18172.

\medskip

7.~A. T. Bajkova, A. A. Smirnov, and V. V. Bobylev. Analysis of regularity/chaoticity of the globular clusters dynamics on the central region of the Milky Way. // Publications of the Pulkovo Observatory {\bf 233}, P. 1--28 (2024) DOI:10.31725/0367-7966-2024-233-1-28,
arXiv: 2406.15590.

\medskip

8.~E. Vasiliev, H. Baumgardt. Gaia EDR3 view on galactic globular clusters. // Monthly Notices of the Royal Astronomical Society {\bf 505}, Issue 4, P. 5978--6002 (2021), arXiv: 2102.09568.

\medskip

9.~H. Baumgardt, E. Vasiliev.  Accurate distances to Galactic globular clusters through a combination of Gaia EDR3, HST, and literature data. // Monthly Notices of the Royal Astronomical Society {\bf 505}, Issue 4, P. 5957--977 (2021), arXiv: 2105.09526.

\medskip

10.~J. Palous, B. Jungwiert, J. Kopecky. Formation of rings in weak bars: inelastic collisions and star formation. // Astronomy and Astrophysics
{\bf 274}, P. 189--202 (1993).

\medskip

11.~J. L. Sanders, L. Smith, N. W. Evans, P. Lucas. Transverse kinematics of the Galactic bar-bulge from VVV and Gaia. // Monthly Notices of the Royal Astronomical Society {\bf 487}, Issue 4, P. 5188--5298 (2019), arXiv: 1903.02008.

\medskip

12.~R. E. G. Machado, T. Manos. Chaotic motion and the evolution of morphological components in a time-dependent model of a barred galaxy within a dark matter halo. // Monthly Notices of the Royal Astronomical Society {\bf 458}, Issue 4, P. 3578--3591 (2016), arXiv: 1603.02294.

\medskip

13.~M. Miyamoto, R. Nagai. Three-dimensional models for the distribution of mass in galaxies. // Publications of the Astronomical Society of Japan  {\bf 27}, P. 533--543 (1975).

\medskip

14.~J. F. Navarro, C. S. Frenk, S. D. M. White. A Universal Density Profile from Hierarchical
Clustering. // Astrophysical Journal {\bf 490}, Issue 2, P. 493=-508 (1997), arXiv: astr-ph/9611107.

\medskip

15.~A. T. Bajkova, V. V. Bobylev. Rotation curve and mass distribution in the Galaxy from the
velocities of objects at distances up to 200 kpc. // Astronomy Letters {\bf 42}, Issue 9, P. 567--582 (2016), arXiv: 1607.08050.

\medskip

16.~P. Bhattacharjee, S. Chaudhury, S. Kundu. Rotation Curve of the Milky Way out to ~200
kpc. // Astrophysical Journal {\bf 785}, Issue 1, id.63, 13 pp. (2014), arXiv: 1310.2659.

\medskip

17.~A. Bajkova, V. Bobylev. Parameters of Six Selected Galactic Potential Models. // Open Astronomy {\bf 26}, Issue 1, P. 72--79 (2017), arXiv: 1801.08875.

\medskip

18.~R. Sch\"onrich, J. Binney, W. Dehnen. Local kinematics and the local standard of rest. // Monthly Notices of the Royal Astronomical Society {\bf 403}, Issue 4, P. 1829--1833 (2010), arXiv: 0912.3693.

\medskip

19.~V. V. Bobylev, A. T. Bajkova. Analysis of the Z distribution of young objects in the Galactic thin disk. // Astronomy Letters {\bf 42}, Issue 1, P. 1--9 (2016), arXiv: 1511.08438.

\medskip

20.~E. Vasiliev. Proper motions and dynamics of the Milky Way globular cluster system from Gaia
DR2. // Monthly Notices of the Royal Astronomical Society {\bf 484}, Issue 2, P. 2832--2850 (2019), arXiv: 1807.09775.

\medskip

21.~D. Massari, H. H. Koppelman, A. Helmi. Origin of the system of globular clusters in the Milky
Way. // Astronomy and Astrophysics {\bf 630}, P. L4, 8 pp. (2019), arXiv: 1906.08271.

\medskip

22.~A. T. Bajkova, G. Carraro, V. I. Korchagin, N. O. Budanova, V. V. Bobylev. Milky Way
Subsystems from Globular Cluster Kinematics Using Gaia DR2 and HST Data. // Astrophysical Journal {\bf 895}, Issue 1, id.69, 18 pp. (2020), arXiv: 2004.13597.

\medskip

23.~A. Morbidelli. // Modern Celestial Mechanics. Aspects of Solar System Dynamics. London and New York, 432 pp. (2014).

\medskip

24.~N. Nieuwmunster, M. Schultheis, M. Sormani, F. Fragkoudi, F. Nogueras-Lara, R. Schodel, P. McMillan. Orbital analysis of stars in the nuclear stellar disc of the Milky Way. // Astronomy \& Astrophysics {\bf 685}, id.A93, 19 pp. (2024), arXiv: 2403.00761.

\medskip

25.~M. Valluri, V. P. Debattista, T. Quinn, B. Moore. The orbital evolution induced by barionic condensation in triaxial halos. // Monthly Notices of the Royal Astronomical Society {\bf 403}, Issue 1, P. 525--544 (2010), arXiv: 0906.4784.
}

 \end{document}